\documentclass[runningheads]{llncs}
\usepackage[T1]{fontenc}
\usepackage{array}
\usepackage{makecell}
\usepackage{longtable}
\usepackage{booktabs}
\usepackage{tabularx} 
\usepackage{tabularx}
\usepackage{array}
\usepackage{float} 
\usepackage{hyperref} 
\usepackage{tabularx}  
\usepackage{booktabs}  
\usepackage{caption}   

\usepackage{tcolorbox}

\usepackage{hyperref}

\usepackage{xcolor}
\usepackage{graphicx}
\begin{document}
\title{From Specification to Service: Accelerating API-First Development Using Multi-Agent Systems}
%
%
\author{
  Saurabh Chauhan\inst{1} \and
  Zeeshan Rasheed\inst{1} \and
  Malik Abdul Sami\inst{1} \and
  Kai-Kristian Kemell\inst{1} \and
  Muhammad Waseem\inst{1} \and
  Zheying Zhang\inst{1} \and
  Jussi Rasku\inst{1} \and
  Mika Saari\inst{1} \and
  Pekka Abrahamsson\inst{1}
}

\titlerunning{Accelerating API-First Development with Multi-Agent Systems }
\authorrunning{S. Chauhan et al.}
\institute{Faculty of Information Technology and Communication Science,  
  Tampere University, Finland\\
\email{\{saurabh.chauhan, zeeshan.rasheed, malik.sami, muhammad.waseem, kai-kristian.kemell, zheying.zhang, jussi.rasku, mika.saari, pekka.abrahamsson\}@tuni.fi} \\
\url{https://www.tuni.fi/en}
}

\maketitle              
\begin{abstract}
This paper presents a system that uses Large Language Models (LLMs) based agents to automate the API-first development of RESTful microservices. This system helps to create an OpenAPI specification, generate server code from it, and refine the code through a feedback loop that analyzes execution logs and error messages. The integration of log analysis enables the LLM to detect and address issues efficiently, reducing the number of iterations required to produce functional and robust services. This study's main goal is to advance API-first development automation for RESTful web services and test the capability of LLM-based multi-agent systems in supporting the API-first development approach. 
To test the proposed system's potential, we utilized the PRAB benchmark. The results indicate that if we keep the OpenAPI specification small and focused, LLMs are capable of generating complete functional code with business logic that align to the specification. 
The code for the system is publicly available at \url{https://github.com/sirbh/code-gen}.

\end{abstract}
\section{Introduction}
\label{Introduction} 

Large Language Models (LLMs) have transformed various domains \cite{naveed2023comprehensive}, showed capabilities in understanding, generating, and processing human language \cite{chang2024survey}. Their advanced reasoning and generative abilities are increasingly being leveraged in software engineering, particularly for automating complex development tasks \cite{fan2023large}. However, LLMs often face difficulties in achieving accurate and effective problem-solving, especially when the task requires a high degree of meaningful collaboration \cite{hong2023metagpt}. To this end, LLM-based agents, designed to perform specific roles and interact autonomously, are emerging as a promising paradigm for enhancing developer productivity and optimizing software creation workflows \cite{li2024survey}.

The API-first approach is a widely adopted way of building microservices, which promotes the design of APIs using standard tools such as OpenAPI specification, before their implementation to ensure consistency, reusability, and efficient integration \cite{enase25} \cite{lercher2023microservice}. However, traditional methods often require manual intervention for writing specifications for the services, switching between different tools for debugging, error resolution, to ensure the functional correctness of generated services, leading to long development cycles and increased resource expenditure \cite{lazar2024specrawler} \cite{enase25}.

To address these challenges, we proposed an LLM-based multi-agent system to automate the entire API-first development lifecycle for RESTful microservices. A core innovation of the proposed system is to run the code locally, which ensures that the generated code functions as expected in the user's specific setup, addressing environment-specific issues right away. It also provides feedback based on logs generated and suggests fixes based on those logs, which helps the user resolve issues quickly. 
In this study, we extend the findings of our previous work \cite{enase25} by using the PRAB benchmark to expand our evaluation process. The results indicate that the system successfully generated all specifications selected from the PRAB benchmark based on natural language prompts. Furthermore, for all the considered specifications, the system was able to generate code that accurately aligns with the respective API definitions. In addition, we have developed an automated testing pipeline for the specification generation component.

Our contributions can be summarized as follows:
\begin{itemize}
    \item We propose and implement LLM-based multi-agent system designed for the autonomous API-first development of RESTful microservices.

    \item We integrate a feedback loop mechanism that leverages execution log analysis and error message interpretation to enable correction and refinement of generated code.

    \item We evaluated the proposed system using the filtered PRAB benchmark to test the system's capabilities and effectiveness.

    \item 
    The proposed system's code and benchmark evaluation results are publicly available for further validation.
\end{itemize}

\section{Background}
\label{Background}

\subsection{Generative AI}
\label{Generative AI}

Generative Artificial Intelligence (AI) refers to a class of machine learning models that can generate new and intricate data, such as images or text, by learning from the patterns present in pre-existing data \cite{10421601}. The influence of Generative AI can be observed across various domains, such as technology, business, education, healthcare, and the arts \cite{10628898}, \cite{10589972}, \cite{rasheed2024can}. However, it introduces many challenges such as evaluation difficulties, ethical issues, and data quality. To this end, several researchers suggested that there is a need for proper AI-human alliance so that these challenges can be eliminated and benefits can be maximized \cite{10556223}, \cite{9889652}, \cite{10421601}, \cite{rasheed2024timeless}. Furthermore, this technology also presents several other technical challenges. For instance, hallucination is a common issue, where the AI produces content that appears plausible but is factually incorrect or fabricated \cite{alkaissi2023artificial}. This can lead to serious consequences depending on the industry \cite{sallam2023chatgpt}.

One way to address concerns regarding the accuracy of AI-generated content is by incorporating Human-in-the-Loop (HITL) approaches. By involving humans in tasks where precision is critical, the risk of hallucinations or incorrect outputs can be significantly reduced \cite{kathiresan2025human}. Additionally, human feedback provides rapid evaluations of the generated content, helping the model to refine and improve its output accuracy over time \cite{christiano2017deep}.

\subsection{Large Language Models for Code Generation}
\label{GPT models in SE} 
In recent years, LLMs have become a popular choice for code generation tasks. These models combine natural language understanding with generative capabilities, leading to exceptional performance in code synthesis \cite {liu2023codegeneratedchatgptreally}, \cite{rasheed2024ai}. This has attracted the attention of many academic researchers and software developers \cite{jiang2024survey}. Another notable application of these models is code completion, which suggests code snippets based on partially written code.

A more recent development in generative AI for code-related tasks is the integration of function-calling capabilities into LLMs. Function calling enables models to execute structured API requests, interact with external systems, and automate complex workflows \cite{gim2024asynchronous}. Rather than just generating code as output, these LLMs can invoke predefined functions, retrieve real-time data, and even perform code execution \cite{kim2024llmcompilerparallelfunction}. This feature is particularly valuable in software development, as it facilitates seamless integration with APIs, debugging tools, and deployment environments \cite{gim2024asynchronous}.

Despite their many benefits, LLMs in code generation tasks face several challenges. One of the most common and concerning is the quality of the generated code, which often contains bugs or security vulnerabilities. As a result, human involvement in the process of using LLMs to solve a particular problem becomes extremely crucial \cite{kathiresan2025human} \cite{abdelghafour2024hallucination}. Furthermore, fine-tuning the model on datasets containing vulnerability fixes can also address the security concerns in generated code \cite{wang2023enhancinglargelanguagemodels}.

\subsection{ API-First Approach in Microservices}
\label{API-First Approach}

\begin{figure}
    \centering
    \includegraphics[width=1\linewidth]{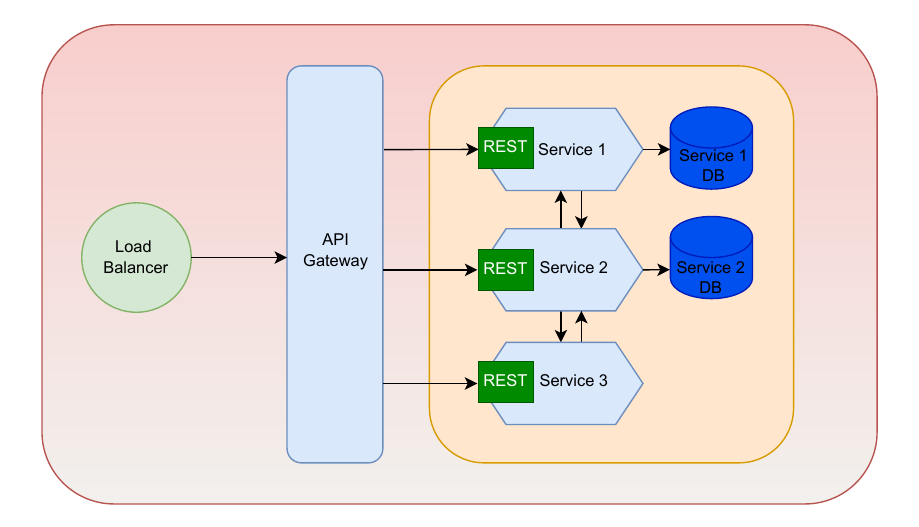}
    \caption{Microservice architecture \cite{enase25}}
    \label{fig:microservice}
\end{figure}

An API-first approach is a development methodology that prioritizes the design and formal definition of Application Programming Interfaces (APIs) before any of the back-end implementation begins \cite{beaulieu2022api}. The API's inputs and outputs are initially discussed and defined in a concrete manner, ensuring a clear understanding of its use cases and communication protocols \cite{dudjak2020api}. This process establishes a shared understanding and clear contracts between different services, which in turn promotes modularity since both API clients and suppliers are only required to adhere to these contractual rules. Essentially, this means that API definitions are treated as first-class citizens \cite{dudjak2020api}.

In microservice architecture, the system is first broken down into smaller independent services. Then API contracts of these independent services must be defined before the development team starts implementing them \cite{DudjakMario2020AAmf}. Once the interface specification has been established for each service, it is shared among the team so that consumers and providers can work in parallel. This approach promotes faster delivery and enhances reusability. Depending on the type of communication (i.e., synchronous or asynchronous), suitable tools must be selected for writing the specification \cite{DudjakMario2020AAmf}.

To define these API contracts, standards must be established to ensure clarity and consistency, which in turn enhances collaboration. To achieve this standardization, one can use the OpenAPI specification  to define the APIs \cite{9650408}. The OpenAPI specification is a standard format for describing RESTful APIs, which makes them machine-readable and easy to share among teams and systems. It offers a comprehensive definition of an API, including its endpoints, operations, request/response formats, and security requirements. Furthermore, an OpenAPI specifications definition can be used to generate interactive documentation, which allows users to interact with the API and understand its expected responses \cite{9650408}.

Additionally, these specification documents can be versioned to track the changes, speed up the clear communication about the updates done in the API, across various development teams or consumers.

As previously mentioned, the API-first approach treats APIs as a single source of truth. This allows development teams writing code for clients and servers to work in parallel, since a clear and well-defined contract has already been established \cite{DudjakMario2020AAmf}. This concurrent development also extends to testing, enabling testers to plan test cases and validate both client and server implementations more quickly. This synchronized development process allows teams to detect faults early, which leads to rapid iteration and promotes an agile development life cycle \cite{alma9911478695505973}.  

\section{Research Methodology}
In this section, we present the methodology for automating the entire development of
a service from generating the Open API specification to testing and fixing the API code. Section \ref{RQs} provides details of the formulated Research Questions (RQs). The system design and multi-agent workflow are discussed in Section \ref{System Design}, and we present the details of our evaluation framework in Section \ref{Evaluation Framework}.

\subsection{Research Questions (RQs)}
\label{RQs}
Based on our study goal, we formulated the following two Research Questions (RQs).

\begin{tcolorbox}[colback=white!2!white,colframe=black!75!black]
\textit{\textbf{RQ1.} To what extent are LLM-based multi-agent systems capable of autonomously executing the entire OpenAPI-first development process?}
\end{tcolorbox}

The objective of \textbf{RQ1} is to explore how LLM-based agents can be integrated into a system to automate the OpenAPI-first development workflow. This includes generating OpenAPI specifications from input requirements, producing corresponding server code, and incorporating refinement steps using execution feedback.

\begin{tcolorbox}[colback=white!2!white,colframe=black!75!black]
\textit{\textbf{RQ2.} How reliable are the results generated by the proposed system?}
\end{tcolorbox}

The goal of \textbf{RQ2} is to evaluate the accuracy and effectiveness of the outputs generated by the proposed system. This involves assessing how well the system produces valid and complete OpenAPI specifications and server code. The evaluation is conducted using benchmark datasets to measure performance, reliability, and overall correctness.

\subsection{System Design}
\label{System Design}

The main goal of the proposed system is the development of services that perform Create, Read, Update, and Delete (CRUD) operations and will communicate using REST architectural style, which is appropriate for these services because of its ability to handle CRUD operations using HTTP methods such as POST, GET, PUT and DELETE \cite{9320801}.
The proposed system will use the OpenAPI specification standard to define the API specification as it provides a complete framework for defining REST APIs \cite{9650408}.
The LLM that is responsible for interacting with users is GPT-4o by OpenAI because of its great performance in generating accurate code snippets and interpreting complex natural language prompts. Moreover, its function-calling feature makes it a suitable model for this type of system \cite{10298721}. The system uses this function-calling feature of LLM to interact with the user's environment and execute appropriate commands.

The generated API code will be in JavaScript, utilizing the Express.js framework—a lightweight and flexible tool well-suited for constructing RESTful web services. Unlike conventional code generators that mainly produce boilerplate or a basic API skeleton, our system creates fully functional API code, including the necessary business logic. An example of a generated OpenAPI specification and its corresponding code is provided here:  
\url{https://github.com/sirbh/sample_generated_cpi}.
 
\begin{figure*}[t]
    \centering
    \includegraphics[width=0.8\textwidth]{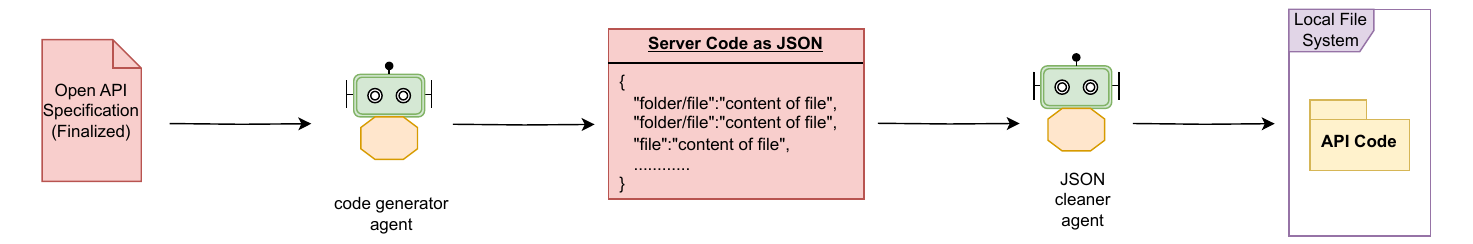} 
    \caption{Server code generation process\cite{enase25}}
    \label{fig:code_generation}
\end{figure*}

\begin{figure*}[t]
    \centering
    \includegraphics[width=0.8\textwidth]{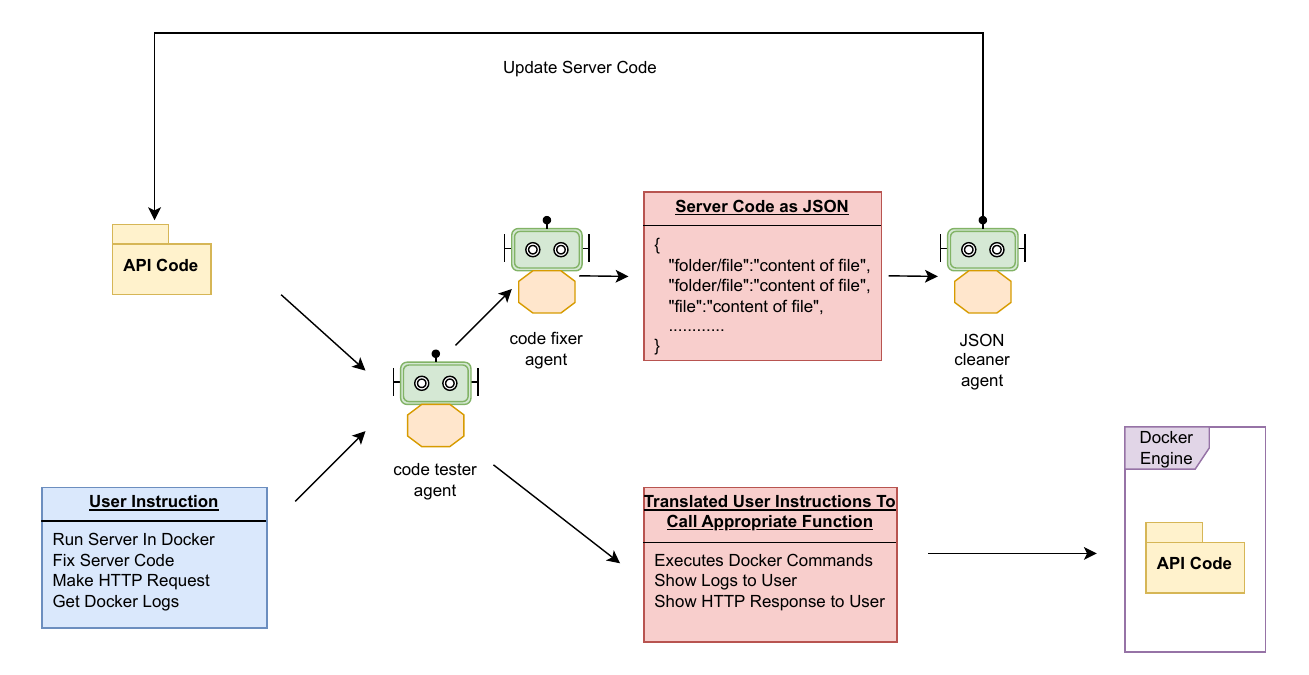} 
    \caption{Interacting with server \cite{enase25}}
    \label{fig:server_interaction}
\end{figure*}

The system's structured workflow comprises three stages, as shown in Figures \ref{fig:spec_generation}, \ref{fig:code_generation}, and \ref{fig:server_interaction}. The process begins with user input and concludes with the testing of the generated server code.
For this task, we employed an LLM-based multi-agent architecture where each agent has a defined role. Details about the agents and the functions they can call to interact with the local environment are provided in Table \ref{tab:generation-result}.

The multi-agent architecture was adopted to enhance the system's modularity, flexibility, and scalability. By dividing the system's functionality into various agents, with each responsible for a specific task like code generation, testing, and specification generation, the system can operate more efficiently and be more easily maintained.

Furthermore, this architecture allows for the replacement or updating of individual agents as needed. If a new language or framework is required, the code-generator agent can be replaced with one specifically designed for that environment. This eliminates the need to rework the entire system, making it easier to update the technology stack without causing disruption.

The system's multi-agent architecture facilitates efficient memory management by isolating responsibilities. Each agent is responsible for its own context; for example, the spec generator agent is exclusively concerned with producing the OpenAPI specification and has no need to retain information about subsequent stages like code generation, debugging, or testing. This division reduces redundant data storage, mitigates potential conflicts, and enhances overall performance by ensuring that each agent processes only the information pertinent to its specific task.

The following are the steps involved in the process of generating, validating, and fixing the code for a service:

\begin{figure*}[t]
    \centering
    \includegraphics[width=0.8\textwidth]{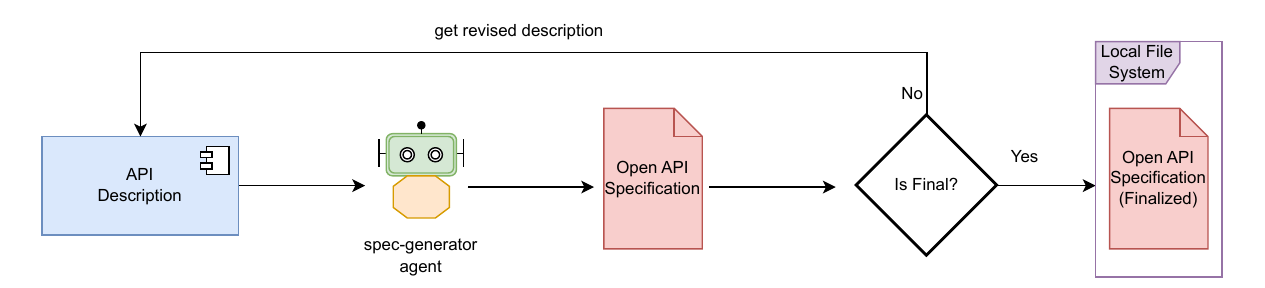} 
    \caption{Specification generation process \cite{enase25}}
    \label{fig:spec_generation}
\end{figure*}

\subsubsection{OpenAPI Specification Generation:}

First, as shown in Figure \ref{fig:spec_generation}, the user provides high-level service requirements. These requirements typically detail the necessary data models, endpoints, and database operations. For instance, a user might provide a prompt such as: "Generate an OpenAPI specification for a product service that can add, edit, delete, and fetch products. A product should have fields for name, description, price, and quantity." Using this input, the system leverages a GPT-4o model, handled by the spec-generator agent, to produce the initial version of the OpenAPI specification. This is a crucial step for ensuring that the service interface is clearly defined and align to OpenAPI standards. An example of a generated specification is provided here \url{https://github.com/sirbh/sample_generated_cpi/blob/main/openapi_spec.yml}

\subsubsection{Finalization of specification:}

After the initial OpenAPI specification is generated, it is presented to the user for review and refinement. Users can provide feedback and make adjustments to the spec-generator agent in an iterative process that continues until the specification accurately reflects their requirements. Once the user is satisfied with the output, they can indicate that the specification is final, as shown in Figure \ref{fig:spec_generation}. Using GPT-4o's function-calling feature, the agent then invokes a function that takes the specification in string format and saves it to the user's local environment. This finalized version serves as a blueprint or contract for the rest of the service's development, ensuring consistency and accuracy throughout the process.

\subsubsection{Server Code Generation:}

The finalized OpenAPI specification is then passed to the Server Code Generator, which comprises two agents: the code-generator and the JSON-cleaner, as illustrated in Figure \ref{fig:code_generation}.
The code-generator agent receives the specification and, guided by prompts detailing the desired folder structure, programming language, and framework, outputs a JSON string. This JSON is structured to represent the server's file system, where keys correspond to file paths (e.g., server/index.js) and the associated values are the file's content.

Since this JSON string may contain invalid tokens that could cause parsing errors, the system forwards it to the JSON-cleaner agent. This agent sanitizes the string, allowing it to be parsed without runtime errors. The JSON-cleaner then calls a function that parses the cleaned JSON, creates the necessary directories and files, and populates them with the corresponding content.

The outcome of this step is a complete and logically structured server codebase stored in the user's local environment. This includes all required files, such as the implemented server and business logic, a database service, and configurations necessary for a Docker environment. An example of the generated server code is available here \url{https://github.com/sirbh/sample_generated_cpi/tree/main/express-server}.

\begin{table*}[ht]
\centering
\caption{Agents details \cite{enase25}} 
\label{tab:generation-result}
\resizebox{\textwidth}{!}{%
\begin{tabular}{l|l|l|l}
\hline
\textbf{Agent} & \textbf{Job}                                                                                                                                                        & \textbf{Function Calls}                                                                                                                                                                                                                                                                                                                                                                                                               & \textbf{Function Description}                                                                                                                                                                                                                                                                                              \\ \hline
spec-generator & \begin{tabular}[c]{@{}l@{}}Generate OpenAPI specification from \\ user API description\end{tabular}                                                                 & \texttt{save\_openapi\_spec}                                                                                                                                                                                                                                                                                                                                                                                             & \begin{tabular}[c]{@{}l@{}}Saves the given OpenAPI specification \\ text to a YAML file and returns success/error.\end{tabular}                                                                                                                                                                                            \\ \hline
code-generator & \begin{tabular}[c]{@{}l@{}}Generate code for server in JSON format \\ and save it in user’s working directory\end{tabular}                                          & \texttt{save\_json}                                                                                                                                                                                                                                                                                                                                                                                                      & \begin{tabular}[c]{@{}l@{}}Validates and fixes a given JSON object \\ before saving it as server code.\end{tabular}                                                                                                                                                                                                        \\ \hline
json-cleaner   & \begin{tabular}[c]{@{}l@{}}Cleans JSON data of server files so that it \\ can be parsed without error\end{tabular}                                                  & --                                                                                                                                                                                                                                                                                                                                                                                                                                    & --                                                                                                                                                                                                                                                                                                                         \\ \hline
code-fixer     & \begin{tabular}[c]{@{}l@{}}Takes server code in JSON format with \\ instructions for making fixes or updates \\ and updates the code\end{tabular}                   & \texttt{save\_json}                                                                                                                                                                                                                                                                                                                                                                                                      & \begin{tabular}[c]{@{}l@{}}Validates and fixes a given JSON object \\ before saving it as server code.\end{tabular}                                                                                                                                                                                                        \\ \hline
code-tester    & \begin{tabular}[c]{@{}l@{}}Executes Docker commands to start \\ containers and fetch logs. Sends HTTP \\ requests to services and updates server code.\end{tabular} & \begin{tabular}[c]{@{}l@{}}\texttt{run\_docker\_compose} \\ \texttt{check\_status} \\ \texttt{get\_docker\_logs} \\ \texttt{run\_curl\_command} \\ \texttt{update\_json}\end{tabular} & \begin{tabular}[c]{@{}l@{}}Start Docker containers \\ Check container status \\ Get container logs \\ Send HTTP requests \\ Fix code based on logs \end{tabular} \\ \hline
\end{tabular}
}
\end{table*}

\subsubsection{Automated Validation and Execution:}
\label{Validation}
In order to interact with server code the system uses a code-tester agent. This agent uses the function calling feature of GPT-4o. The most important function it calls helps execute the ``docker-compose up.'' command that builds and loads the docker container in the user's local environment. For example user can give a prompt e.g. \textit{run docker containers} and the agent can use this prompt to match the most appropriate function out of the list of functions that are provided to this agent as shown in Table \ref{tab:generation-result}. It can then call the function that will execute \textit{docker compose up --build}.

Also, it can call functions that can help in fetching the logs of the container and the status of all the services running related to the server code if the prompts e.g.\textit{get logs related to service} or \textit{get service status} are provided. After calling the appropriate function it uses the returned data to show a summary of what's happening in the docker engine in a readable format. It eliminates the need to read the console logs which are often not user-friendly. 

Moreover, to validate the working of the server, the user can use natural language prompts e.g. \textit{get the list of products} or \textit{delete product with id}. The agent will then make requests to the server container by calling the function and getting the output to the user which eliminates the need to switch to any other tool like Postman to make requests and validate functionality. 

\subsubsection{Iterative Code Fixing:}

After the user can interact with server code i.e. start and run containers, make requests to server containers, and get logs of running containers, there might be a scenario where the user encounters issues or something does not work as expected. In such cases, the code-tester agent, which has access to the server logs through its memory context if the user has asked for the logs (as discussed in previous steps), can assist. If the user asks the agent to detect the problem, the agent can analyze the logs and suggest potential fixes. 

Furthermore, it can interact with the code-fixer agent as shown in Figure \ref{fig:server_interaction} to modify the already saved server code by doing those fixes and restarting the services again to run the updated code. This eliminates the requirement of users to look through the code and make the updates, hence it can help in increasing productivity and saving time. To achieve this, the code tester agent prompts the code-fixer with the issue in code and server code in JSON format. The code-fixer agent then calls a function that takes two inputs i.e. the server code and the potential fix to a encountered problem. 

The output of the code-fixer is again a JSON string with keys as directories and values as the content of the files. This code is again parsed and saved, which updates the server code on the user's working environment. After the code is updated, the user can give a prompt to the code tester agent to rebuild and restart all the services and validate the server code, if it's working as expected. This functionality is repeated by keeping the user in the loop until the desired result is achieved.

\subsection{Evaluation Framework}
\label{Evaluation Framework}
In this section, we present the evaluation framework used to test the system. The objective was to verify the correctness of components such as as specification generation, code generation, and interaction with the generated code.

Section~\ref{sec:benchmark-select} introduces the benchmark used in our evaluation. Section~\ref{sec:openAPI specification eval framework} describes the framework for evaluating the specification generation component. Section~\ref{sec:Code Generation Eval Framework} details the methodology for assessing code generation. Finally, Section~\ref{sec:Runtime Interaction and Validation eval framework} presents the framework used to evaluate the system's ability to interact with and validate the generated code during runtime.

\subsubsection{PRAB Benchmark:}
\label{sec:benchmark-select}

To evaluate the OpenAPI specification generation and code generation capabilities of our system, we required a set of real-world and diverse API specifications. For this purpose, we selected specifications from the PRAB (Public REST API Benchmark) benchmark \cite{decrop2025public}, which provides a curated collection of OpenAPI definitions for a wide range of services. PRAB benchmark
contains documentation and structural characteristics of 60
publicly available REST APIs \cite{decrop2025public}.

\begin{table*}[ht]
\centering
\caption{PRAB specifications}
\label{tab:filterd-spec-prab}
\begin{tabular}{l|r|r}
\hline
\textbf{Spec File}                 & \textbf{Lines} & \textbf{Paths} \\ \hline
balldontlie-openapi.json           & 225            & 10             \\ \hline
google-geocoding-openapi.json      & 140            & 1              \\ \hline
open-brewery-db-openapi.json       & 207            & 5              \\ \hline
piggy-metrics-openapi.json         & 241            & 8              \\ \hline
quartz-manager-openapi.json        & 276            & 5              \\ \hline
random-user-generator-openapi.json & 243            & 1              \\ \hline
rest-faults-openapi.json           & 205            & 4              \\ \hline
\end{tabular}
\end{table*}
\paragraph{Data Filtering:}
To test the ability of our multi-agent system, we initially apply a data filtering process to the PRAB dataset. Given that our proposed system operates on a microservices-based architecture, designed to generate code for smaller services, we filtered the PRAB dataset and extracted only those specifications corresponding to files with a maximum of 300 lines of code. These spec are listed in Table \ref{tab:filterd-spec-prab}.

\subsubsection{Evaluation of Specification Generation:}
\label{sec:openAPI specification eval framework}

To evaluate the ability of our multi-agent system to generate accurate OpenAPI specifications from natural language prompts, we used a filtered PRAB dataset. The evaluation focuses on the system's capability to construct valid and semantically equivalent specifications through an iterative process.

\begin{figure}
    \centering
    \includegraphics[width=1\linewidth]{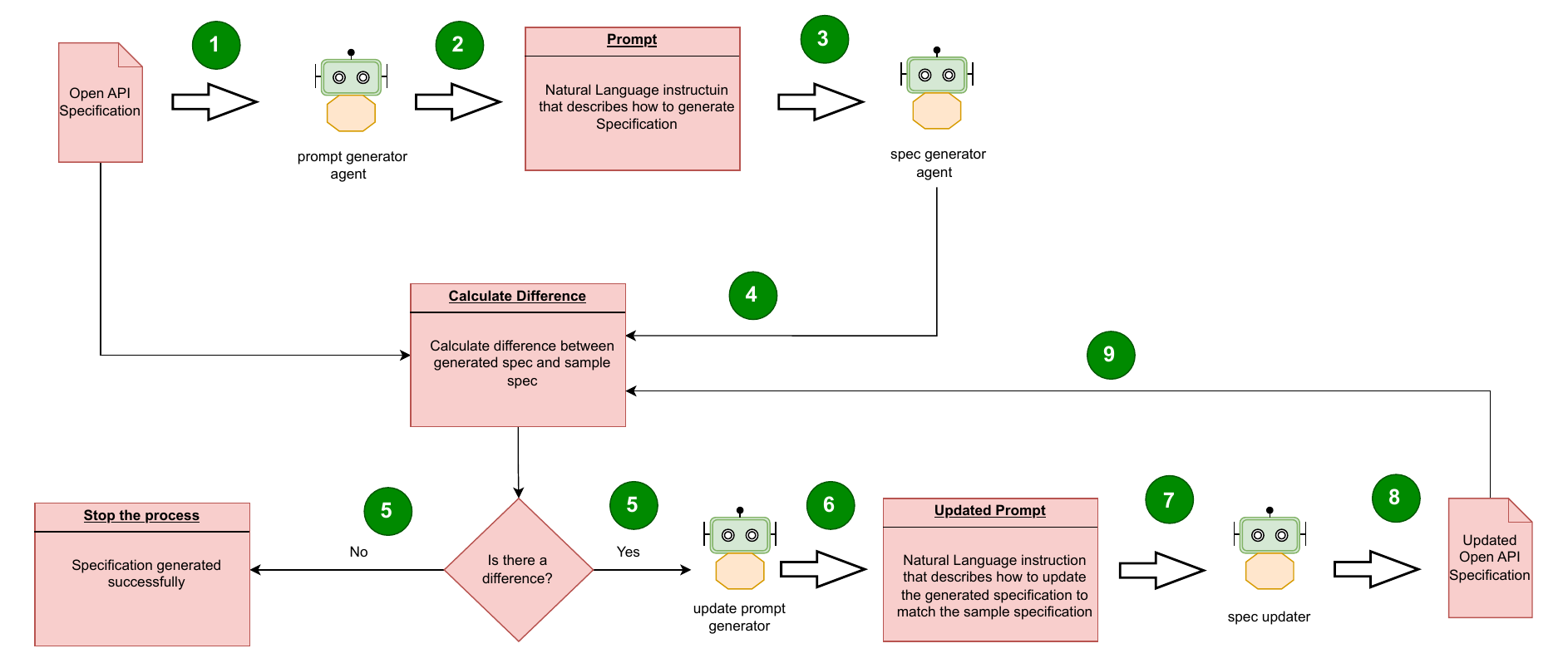}
    \caption{Automated testing setup}
    \label{fig:spec gen test}
\end{figure}

As shown in Figure \ref{fig:spec gen test} we began by selecting a sample of OpenAPI specifications from the filtered PRAB dataset. Each selected OpenAPI specification served as a ground truth reference for the evaluation.
To generate the corresponding natural language prompt for each specification, we used a separate agent. This prompt was then passed to our \textit{specification generator agent}, which produced an initial OpenAPI specification from the natural language instruction.

The generated specification was then compared with the original specification using a structural diff tool \cite{deepdiff}, and the differences between the specifications were captured. 
Then, we provided these differences to another agent, which was prompted to generate detailed instructions in natural language to update the generated specification, aiming to bring it closer to the reference specification. This iterative cycle continued until the generated specification matched the original or no further meaningful improvements could be made. All intermediate prompts, specifications, and differences were logged for analysis. This setup allowed us to quantify how well the system could generate specifications similar to the original specification by simply receiving instructions in natural language.

\subsubsection{Code Generation Evaluation:}
\label{sec:Code Generation Eval Framework}
To evaluate the code generation capabilities of the system, we also used the PRAB dataset. The system was tasked for generating server-side code from the specification. We then evaluated the generated code on multiple criteria. First, we checked that the output followed the expected folder structure provided in the code generator agent prompt \cite{sirbh_code_generator_prompt}. Next, we tested whether the code could be successfully built and executed using a \textit{docker-compose.yml} file generated by the system. For those cases where the generated code resulted in errors or failed to run locally, the faulty code was fed back into the system to get fixed. This allowed us to assess the system’s ability to automatically identify issues and apply corrective changes. The fix-and-retry process was repeated until the server was running and aligned with the original OpenAPI specification. Lastly, we also checked if the generated code has all the endpoints and HTTP methods implemented as mentioned in the specification.

\subsubsection{Runtime Interaction and Validation:}
\label{sec:Runtime Interaction and Validation eval framework}

\begin{figure}[htbp] 
    \centering 
    \begin{tcolorbox}[
        colback=white!95!white, 
        colframe=gray!75!black, 
        boxrule=0.8pt,          
        arc=4pt,                
        left=6pt, right=6pt, top=6pt, bottom=6pt, 
        title=\textbf{User Inputs and Expected Tool Calls}, 
        fonttitle=\bfseries,    
    ]

    \textbf{User Input:} \texttt{"Start the backend services"} \\
    \quad $\Rightarrow$ \textbf{Expected Tool Call:} \texttt{run\_docker\_compose\_up}

    \vspace{6pt}\hrule\vspace{6pt} 

    \textbf{User Input:} \texttt{"Can you show me the logs?"} \\
    \quad $\Rightarrow$ \textbf{Expected Tool Call:} \texttt{get\_all\_docker\_logs}

    \vspace{6pt}\hrule\vspace{6pt} 

    \textbf{User Input:} \texttt{"Please run docker compose and give me the logs"} \\
    \quad $\Rightarrow$ \textbf{Expected Tool Calls:} \texttt{run\_docker\_compose\_up}, \texttt{get\_all\_docker\_logs}

    \vspace{6pt}\hrule\vspace{6pt} 


    \textbf{User Input:} \texttt{"Update the backend server to support user auth"} \\
    \quad $\Rightarrow$ \textbf{Expected Tool Call:} \texttt{update\_server\_code}

    \vspace{6pt}\hrule\vspace{6pt} 

    \textbf{User Input:} \texttt{"Change the server logic to return JSON instead of XML"} \\
    \quad $\Rightarrow$ \textbf{Expected Tool Call:} \texttt{update\_server\_code}

    \end{tcolorbox}
    \caption{Examples of user interactions demonstrating system tool calling} 
    \label{fig:user_tool_interaction_examples} 
\end{figure}

This part of the evaluation focused on testing the system's ability to correctly interact with the generated server code during runtime. This functionality was heavily dependent on the function-calling capabilities of the underlying LLM, which acted as an agent to invoke the appropriate tools (e.g., code update, request execution, container management) based on user prompts. To evaluate this behavior, we created a custom unit test cases where each test case simulated a possible user prompt, and the system’s response was checked to see if the correct tools were invoked. 

The test cases were categorized into three main types: (1) cases where the server code ran without errors and only required container startup, (2) cases that tested the ability of the system to send HTTP requests to specific endpoints, and (3) cases where the server code required fixing, thereby testing the system’s ability to invoke the code repair tool. This categorization helped ensure the system’s decision-making ability across different scenarios.

This test setup enabled us to systematically verify whether the system was able to route natural language instructions to the correct underlying tools, ensuring that the system was able to understand the user's intent.

\section{Results}
\label{preliminary result}

In Section \ref{RQ1: Proposed system}, we present the overall result of the proposed system. Section \ref{RQ2: Evaluation Result} reports the evaluation results of the different system components.

\subsection{Proposed System (RQ1)}
\label{RQ1: Proposed system}
\begin{figure}[h]
    \centering
    \includegraphics[width=1\linewidth]{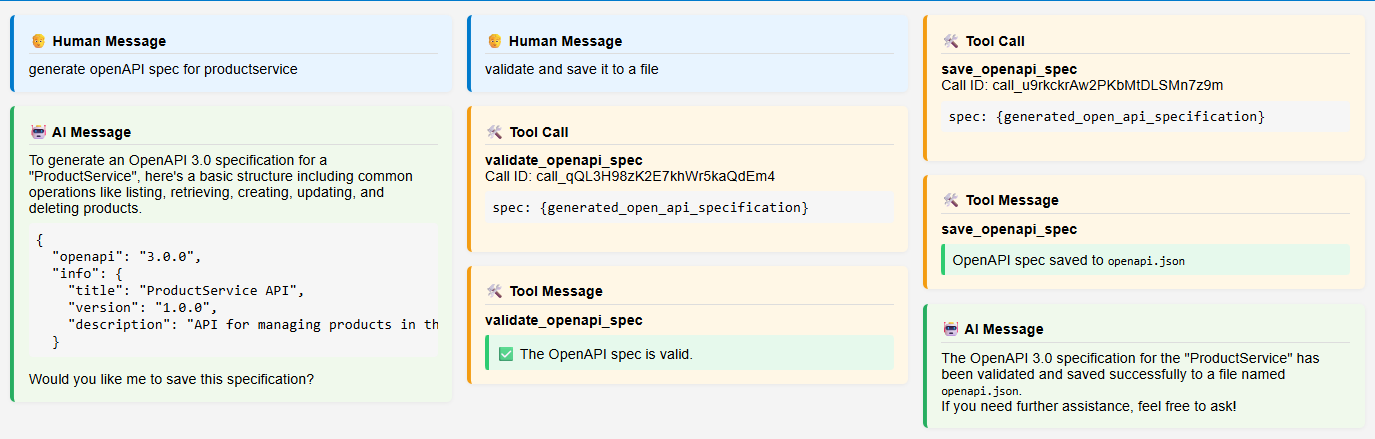}
    \caption{Specification generation chat outcome}
    \label{fig:spec-gen-chat}
\end{figure}

\begin{figure}
    \centering
    \includegraphics[width=1\linewidth]{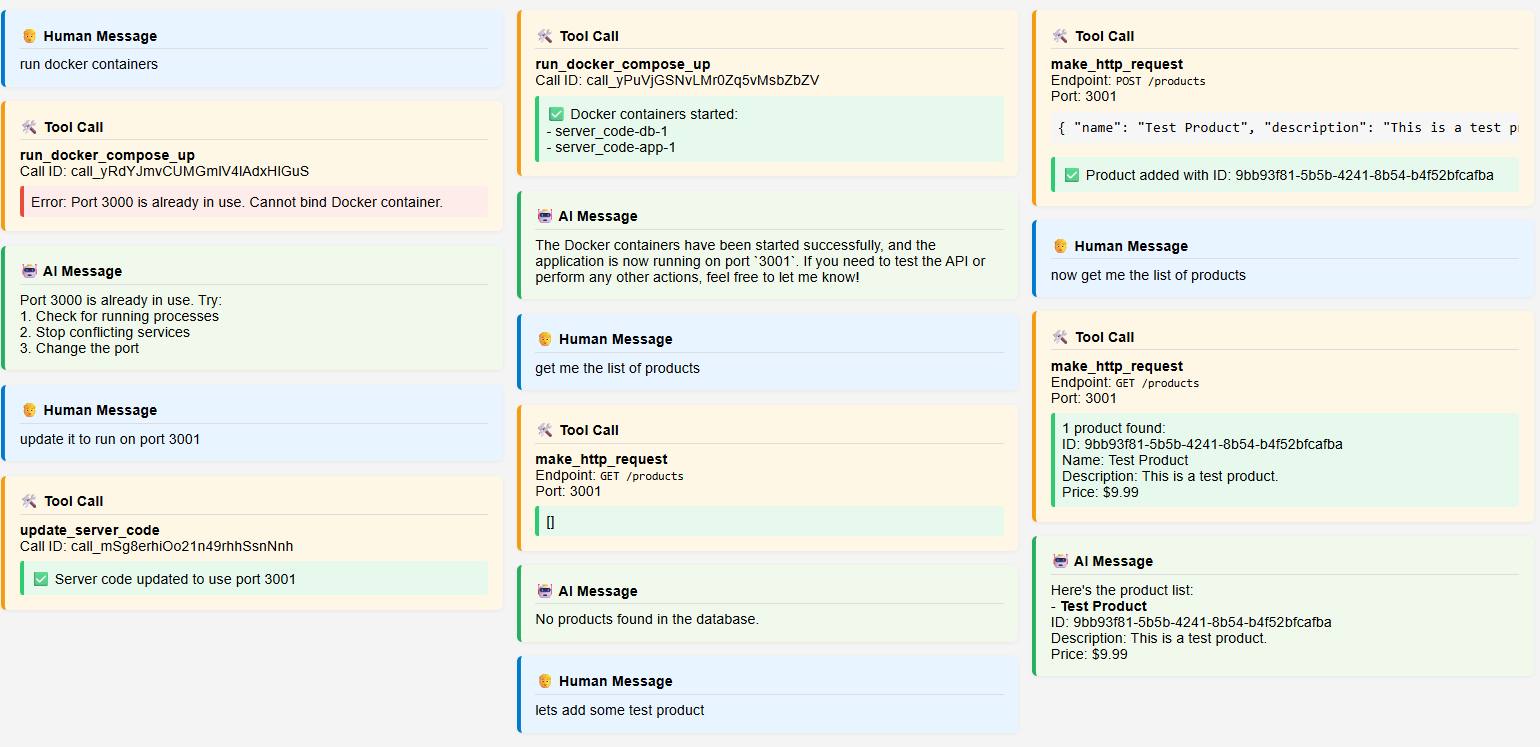}
    \caption{Interacting with server code}
    \label{fig:code-fix-chat}
\end{figure}

The implementation of an LLM-based multi-agent system demonstrates a structured approach to automating the OpenAPI-first development process for RESTful microservices. Initial results indicate that such a system is capable of generating OpenAPI specifications, producing corresponding server code, and refining the output through an iterative feedback loop that incorporates execution logs and error messages. As shown in the Figure \ref{fig:spec-gen-chat}, we were able to generate and validate the specification using natural language prompts using the tool calling feature of LLM. Furthermore, as shown in Figure \ref{fig:code-fix-chat}, we initially prompted the system to run the server containers. It correctly request the necessary tool and identified an issue: \textit{``the target port was already in use''}. The system not only detected the problem but also suggested a potential solution. When instructed to change the port to 3000, the system updated the code accordingly, restarted the containers, and retrieved logs confirming that all services were running normally.

Following this, we prompted the system to make an HTTP request to test the running service. Once again, the system correctly interpreted the instruction, selected the appropriate tool, and supplied the necessary parameters to execute the request successfully. These interactions demonstrate the system’s ability to reason over context and dynamically request the right tools to perform server-side operations in an automated fashion. This entire conversation with more detailed output is available on Github \cite{sirbh_ene2endchat}.

The agents interact within a coordinated workflow to ensure accurate and consistent generation of OpenAPI specifications and related server code. Additionally, a chat-based interface enables developers to interact with the system using natural language prompts. This interface allows for updates and corrections to both specifications and code without requiring manual editing.

The generated code follows to a predefined project structure and is compatible with containerized deployment in Docker environments. Furthermore, the system is equipped to access logs from the local development environment. This capability enables it to support developers in identifying and resolving errors by providing context-aware feedback and suggestions.

By integrating these functionalities into a unified interface, the system reduces the reliance on external tools and manual inspection of logs, error traces, and documentation. Developers can obtain actionable insights and automated fixes directly within the system, facilitating a consistent development workflow.

\subsection{Evaluation Results (RQ2)}
\label{RQ2: Evaluation Result}

In this section, we present the evaluation results for the individual components of the system. First, we report the evaluation results for OpenAPI specification generation, followed by the evaluation results for code generation.

\subsubsection{OpenAPI Specification Generation Evaluation Result:}
\label{OpenAPI Specification Generation Evaluation Result}

To systematically evaluate the OpenAPI specification generation with natural language input, we developed an automated evaluation pipeline as discussed in section \ref{sec:openAPI specification eval framework}. The setup enables recursive refinement of the generated specification, saving version-wise artifacts for later analysis.

\paragraph{Visualization of Spec Refinement:}

To understand how the generated OpenAPI specifications evolved over successive iterations, we visualized the reduction in structural differences across versions. For each specification under test, we computed the number of non-empty lines in the difference files saved during each update iteration. These differences represent the structural or semantic discrepancies between the original input specification and the current generated version.

We plotted the number of diff lines against each version for every specification. Each line in the graph represents a specific OpenAPI spec file, and the y-axis shows the number of differing lines with respect to the original spec, while the x-axis shows the version number of the generated spec. A dip toward zero on the y-axis indicates that the difference between the two specifications was reducing with each iteration.

\begin{figure}
    \centering
    \includegraphics[width=1\linewidth]{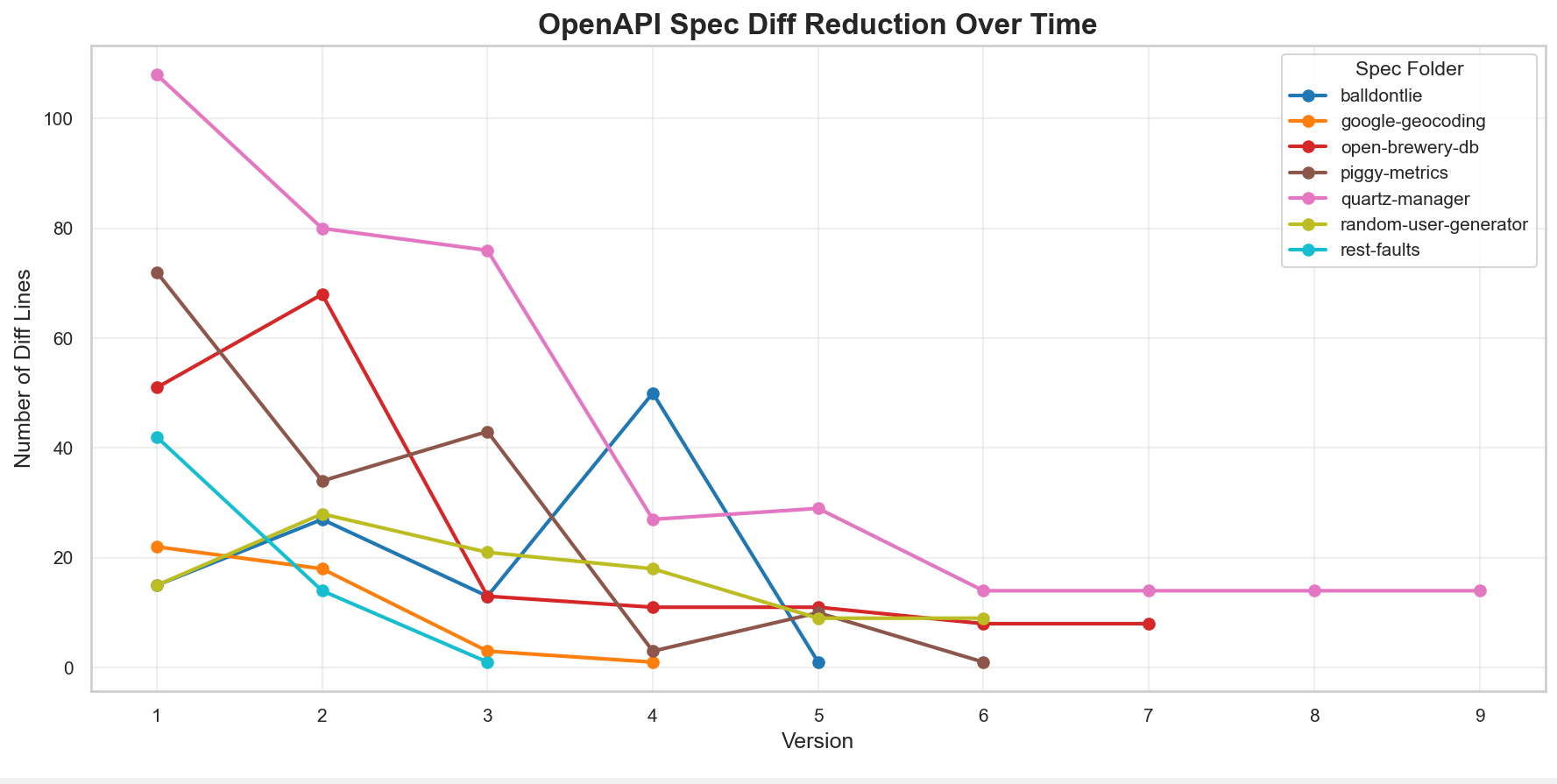}
    \caption{Specification generation test result}
    \label{fig:spec gen visualize}
\end{figure}

As shown in Figure~\ref{fig:spec gen visualize}, many of the lines visibly trend downward, often reaching zero. While two specifications continued to show some differences from the original version, these were very minor, like trivial formatting discrepancies, such as numeric values being represented as strings, which do not affect the semantics or functionality of the specification. This demonstrates system was able to translate a natural language prompt into a correct OpenAPI specification. The complete results of the specification generation evaluation can be accessed via GitHub \cite{sirbh_specgenresults}.

\subsubsection{Code Generation Evaluation Result:}
\label{code eval}
To assess the reliability and correctness of the code generation component, we evaluated the output of the system using a set of real-world OpenAPI specifications sourced from the PRAB benchmark. The evaluation focused on checking whether the generated server code was syntactically correct, executable, and aligned to the expected folder structure. Additionally, we verified whether all endpoints defined in the OpenAPI specification were implemented in the generated code.

\paragraph{Runtime Executability:}
Each generated codebase was provided with a \textit{docker-compose.yml} file and tested for successful container execution using \textit{docker-compose up} command. As shown in Table~\ref{tab:generation-result}, five out of seven projects executed successfully in a containerized environment. The failures were primarily due to errors in the generated schema for the database, such as invalid model definitions or mismatches with the underlying database configuration. This first version of the generated code is present on the Github \cite{sirbh_generatedcodev1}.

\begin{table*}[ht]
\centering
\caption{Generated code features}
\label{tab:generation-result}
\resizebox{\textwidth}{!}{%
\begin{tabular}{l|c|c|c}
\hline
\textbf{Project} & \textbf{Docker Compose} & \textbf{Missing Directories} & \textbf{Implemented All Spec Paths} \\
\hline
balldontlie-openapi & Failed & src/models, src/middlewares & Yes \\
\hline
google-geocoding & Success & src/models, src/middlewares & Yes \\
\hline
open-brewery-db & Success & src/models, src/middlewares & Yes \\
\hline
piggy-metrics & Failed & src/models, src/middlewares & Yes \\
\hline
quartz-manager & Success & src/models, src/middlewares & Yes \\
\hline
random-user-generator & Success & src/models, src/middlewares & Yes \\
\hline
rest-faults & Success & src/models, src/middlewares & Yes \\
\hline
\end{tabular}
}
\end{table*}

\paragraph{Folder Structure Compliance:}
In addition to runtime behavior, we checked whether the generated projects aligned to the desired folder structure conventions typically followed in production-grade backend services. Specifically, we looked for the presence of directories mentioned in the prompt of code generator agent \cite{sirbh_agentprompt}. As shown in Table~\ref{tab:generation-result}, none of the generated projects included all the required directories, although they were explicitly requested in the prompt. This indicates a shortcoming in the system’s ability to fully respect structural instructions when generating boilerplate code.

\paragraph{Alignment With Specification:}

To ensure that the generated server code fully matched the given OpenAPI specification, we manually checked whether all the request paths and HTTP methods were correctly implemented. This involved reviewing the route definitions and controller files in each generated project. We found that the generated code included all the endpoints and methods specified in the OpenAPI specification files. This confirms that the system can reliably translate specifications into working code. The first version of each generated project is available on the GitHub repository at \cite{sirbh_generatedcodev1}.

\paragraph{Code Fixing Capabilities:}

For the two projects where \texttt{docker-compose} initially failed, we provided the generated code back to the system and asked it to fix the issues. The system successfully identified and resolved the problems, and both projects were able to run successfully using \texttt{docker-compose} after the fixes. The corrected versions of both projects are available on GitHub \cite{sirbh_generatedcodev2}. 

Additionally, we have shared the full chat history with the system, which shows how it made appropriate decisions to fix the code. This chat history is also available on the following GitHub \cite{chauhan_chathistoryfixing}.

\begin{tcolorbox}[colback=black!10, colframe=black!80, title=Key Takeaways]
While the system demonstrated a reasonable success rate in generating runnable backend code, its ability to align with expected structural conventions remains limited. These findings suggest improvements in prompts and need to enforce stricter rules for code structure. This could involve enforcing the code generator agent to produce code in a specific format and, if the output does not align with the desired structure, we should provide iterative feedback until the required structure is achieved.
\end{tcolorbox}

\subsubsection{Assessment of Runtime Interaction Capabilities:}
\label{sec:runtime interaction result}
This section presents the evaluation of our system's ability to handle interactions with server code through natural language commands. The evaluation was structured around three key categories of user intents: container management and code update, server code modification to fix errors, and HTTP request execution. 

\begin{itemize}
    \item \textbf{Container Management and Code update}: Commands related to starting, restarting, rebuilding Docker containers or to update something in the code. 
    \item \textbf{Fixing Server Code}: Inputs prompts related to fixing server code for errors.
    \item \textbf{HTTP Request Execution}: Inputs instructing the system to send HTTP requests with various methods (GET, POST, PUT, etc.) to test endpoints.
\end{itemize}

Each user input was expected to trigger one or more tool calls, and the system's response was evaluated to determine whether it could call the correct tool.

\paragraph{Container Management and Code Update Results:}

This section presents the results of unit tests executed to evaluate the system's ability to call tools related to container management and backend code updates. The test cases, listed in the Table \ref{container_management}, include a user input, the expected tool calls based on the intent, and whether the system successfully triggered the correct tool(s).

\begin{table*}[ht]
\centering
\caption{Tool calling test results for container management and code update tasks}
\label{container_management}
\resizebox{\textwidth}{!}{%
\begin{tabular}{l|l|l}
\hline
\textbf{User Input} & \textbf{Expected Tool Calls} & \textbf{Status} \\
\hline
Make sure docker containers are up and running & run\_docker\_compose\_up & success \\
\hline
Start the backend services & run\_docker\_compose\_up & success \\
\hline
Rebuild containers and print the logs & run\_docker\_compose\_up, get\_all\_docker\_logs & success \\
\hline
Update the /api/health endpoint to return a 200 status always & update\_server\_code & success \\
\hline
Just run docker compose up & run\_docker\_compose\_up & success \\
\hline
Include support for email verification in the server code & update\_server\_code & success \\
\hline
Please modify the backend code to log all incoming requests & update\_server\_code & success \\
\hline
Update the backend server to support user authentication & update\_server\_code & success \\
\hline
Restart the services and capture their output logs & run\_docker\_compose\_up, get\_all\_docker\_logs & success \\
\hline
Launch all services from docker compose again & run\_docker\_compose\_up & success \\
\hline
Patch the backend to support file uploads & update\_server\_code & success \\
\hline
Can you show me the logs? & get\_all\_docker\_logs & success \\
\hline
Turn on the backend again & run\_docker\_compose\_up & success \\
\hline
Run the services again, I changed the code & run\_docker\_compose\_up & success \\
\hline
Deploy the updated code and verify the container logs & run\_docker\_compose\_up, get\_all\_docker\_logs & success \\
\hline
Can you rebuild the containers and start the services & run\_docker\_compose\_up & success \\
\hline
Change the server logic to return JSON instead of XML & update\_server\_code & success \\
\hline
Add error tracking in backend for 500 responses & update\_server\_code & success \\
\hline
Please run docker compose and give me the logs & run\_docker\_compose\_up, get\_all\_docker\_logs & success \\
\hline
Start everything and check what's failing & run\_docker\_compose\_up, get\_all\_docker\_logs & success \\
\hline
\end{tabular}
}
\end{table*}

As seen in the Table \ref{container_management}, the system was successful in identifying the correct tools based on user requirement. Since a LLM is involved in interpreting the instructions, results may vary across different runs. For transparency, we have provided a more detailed report including chat history of the agent that illustrates how tool calls were concluded by the agent. More detailed test result are available on the GitHub \cite{chauhan_containermanagement}.

\paragraph{Server Code Fixing Evaluation:}

To evaluate the system’s ability to fix server code, we used a set of inputs representing various user prompts that reflect the need for fixing of code. The test cases and their expected tool calls are shown in the Table \ref{server_fixing}. The evaluation criteria considered whether the correct tools were selected. The test results are also summarized in the Table~\ref{server_fixing}.

\begin{table*}[ht]
\centering
\caption{Tool calling test results for server code fixing tasks}
\label{server_fixing}
\resizebox{\textwidth}{!}{%
\begin{tabular}{l|l|l}
\hline
\textbf{User Input} & \textbf{Expected Tool Calls} & \textbf{Status} \\
\hline
Nothing is being served on localhost. Please investigate and fix. & run\_docker\_compose\_up, get\_all\_docker\_logs, fix\_server\_code & Success \\
\hline
Fix backend code – server is not accepting any connections & fix\_server\_code & Success \\
\hline
Logs show port 8000 is already taken. Resolve the issue. & get\_all\_docker\_logs, fix\_server\_code & Success \\
\hline
The service exits with status code 137 – investigate and resolve & get\_all\_docker\_logs, fix\_server\_code & Success \\
\hline
Services won't start. Find the issue and fix it & fix\_server\_code & Success \\
\hline
Server crashes with memory error, can you fix? & fix\_server\_code & Success \\
\hline
API requests are timing out. Fix the server issue. & fix\_server\_code & Success \\
\hline
Start everything and check what's failing and fix it & run\_docker\_compose\_up, get\_all\_docker\_logs, fix\_server\_code & Success \\
\hline
One of the services fails healthcheck. Fix and restart & get\_all\_docker\_logs, fix\_server\_code & \textbf{Fail} \\
\hline
Docker failed due to address in use. Please check and fix & fix\_server\_code & Success \\
\hline
The health check endpoint is not returning 200 & get\_all\_docker\_logs, fix\_server\_code & Success \\
\hline
The backend isn’t responding after docker compose up & run\_docker\_compose\_up, get\_all\_docker\_logs, fix\_server\_code & Success \\
\hline
The backend throws a 500 error after deployment & fix\_server\_code & \textbf{Fail} \\
\hline
Something is crashing the server on startup, fix it & fix\_server\_code & Success \\
\hline
Fix the error in container startup logs and restart services & get\_all\_docker\_logs, fix\_server\_code & Success \\
\hline
Check if docker is running, and solve any issues found & run\_docker\_compose\_up, get\_all\_docker\_logs & Success \\
\hline
Rebuild the containers, logs are unclear, but fix the root cause & run\_docker\_compose\_up, get\_all\_docker\_logs, fix\_server\_code & Success \\
\hline
I see 'connection refused' in the logs, please fix it & get\_all\_docker\_logs, fix\_server\_code & Success \\
\hline
There's a network error when trying to access the service & get\_all\_docker\_logs, fix\_server\_code & Success \\
\hline
Containers keep restarting in a loop. What’s wrong? & get\_all\_docker\_logs, fix\_server\_code & Success \\
\hline
\end{tabular}
}
\end{table*}

\begin{table*}[ht]
\centering
\caption{Tool calling test results for HTTP request execution tasks}
\label{http_request_execution}
\resizebox{\textwidth}{!}{%
\begin{tabular}{l|l|l}
\hline
\textbf{User Input} & \textbf{Expected Tool Call} & \textbf{Status} \\
\hline
Use DELETE method on /delete-example with query id & make\_http\_request & success \\
\hline
Patch the field in /patch-example with new value & make\_http\_request & success \\
\hline
DELETE the resource with id 42 at /delete-example & make\_http\_request & success \\
\hline
Use PATCH on /patch-example to update a field & make\_http\_request & success \\
\hline
Send a GET request to /get-example & make\_http\_request & success \\
\hline
Send a HEAD request to /head-example & make\_http\_request & success \\
\hline
PUT request to /put-example to change status & make\_http\_request & success \\
\hline
Use HEAD method for /head-example & make\_http\_request & success \\
\hline
Post name and email to /post-example & make\_http\_request & success \\
\hline
Delete using /delete-example with id=10 & make\_http\_request & success \\
\hline
GET the test response from /get-example again & make\_http\_request & success \\
\hline
Can you PUT this object to /put-example? & make\_http\_request & fail \\
\hline
Use OPTIONS to see allowed methods on /options-example & make\_http\_request & success \\
\hline
HEAD request on /head-example to check headers & make\_http\_request & success \\
\hline
PATCH the /patch-example endpoint with an update & make\_http\_request & success \\
\hline
Try calling /get-example using GET method & make\_http\_request & success \\
\hline
POST some data to /post-example & make\_http\_request & success \\
\hline
Use PUT on /put-example with new id and status & make\_http\_request & success \\
\hline
OPTIONS call to /options-example to inspect methods & make\_http\_request & success \\
\hline
Send a POST request to /post-example with user info & make\_http\_request & fail \\
\hline
Check what methods are allowed at /options-example & make\_http\_request & success \\
\hline
\end{tabular}
}
\end{table*}

As seen in the Table \ref{server_fixing}, the system correctly identified and executed the necessary tools in most scenarios, including complex multi-step recovery actions involving container orchestration and code repair. While two cases resulted in failure. Upon closer evaluation, we found out that the system was stuck in a loop and hadn’t encountered a terminal condition, causing it to throw an error.

Detailed logs with chat transcripts are on GitHub \cite{chauhan_fixingerrors}. These logs illustrate how the system made decisions and help explain its reasoning behind each tool call.

\begin{tcolorbox}[colback=black!10, colframe=black!80, title=Key Takeaways]
The system demonstrated a high level of accuracy across all three categories. It successfully interpreted natural language instructions and mapped them to the correct internal tools, validating the tool selection mechanism. Notably, even multi-intent commands were handled effectively.
\end{tcolorbox}

\paragraph{HTTP Request Execution Result:}

We evaluated the system's ability to trigger the correct tool based on different HTTP request instructions. As shown in Table \ref{http_request_execution} all cases except two, the system successfully identified and invoked the appropriate tool \texttt{make\_http\_request}.

Out of 20 total inputs, the system correctly executed 18 requests. Upon investigating the chat history of system, we found that the failed cases were due to insufficient input details, which led the system to skip triggering the tool. More detailed test results are on Githup \cite{chauhan_httprequests}.

\section{Discussion}

The results of the proposed system indicates that when systems are described using formal specifications—such as the OpenAPI standard for REST APIs, LLMs based agents can generate corresponding code and can assist in generating those formal definitions. Furthermore, these kind of system are also able to interact with the user's local environment, such as running servers, saving files, or validating endpoints through function calling abilities. This highlights the potential of LLM-based agents for the autonomous development of software systems.

The findings in this paper suggest that if we can break down software systems into smaller, independent components—each with clearly defined inputs and outputs using some standard protocols, LLMs can be effectively used to generate code for these individual parts. OpenAPI is one such protocol for defining APIs, but similar standards could be developed for other categories of software components. Once these components are generated, they can be integrated to build larger and more complex software systems. In the future, we plan to expand the system by integrating a supervisor agent that can delegate tasks to the current version of the system for generating code for smaller components, and later integrate those components into fully functional software.

Other than this, traditional AI-assisted code generation tools, such as GitHub Copilot, offer significant improvements in developer productivity by providing intelligent code suggestions \cite{Zhang_2023}. However, these systems operate primarily in a static environment, relying on contextual information available within the codebase but lacking real-time feedback from the execution environment \cite{nguyen2022empirical}. This limitation often leads to inaccurate or incomplete code suggestions, as the model does not have access to runtime errors or deployment issues. In contrast, our system takes a more holistic approach by providing suggestions based on output logs in the runtime environment. 

Despite these advantages, there are potential challenges with the proposed system. The accuracy of error detection and fix suggestions depends on the LLM’s understanding of logs and its ability to interpret deployment-specific issues. Further improvements, such as fine-tuning models on domain-specific logs or integrating reinforcement learning-based corrections, could enhance the system’s reliability.

\section{Conclusions}

In this paper, we present a novel LLM-based multi-agent system designed to automate the API-first development lifecycle for RESTful microservices. The proposed system autonomously generates OpenAPI specifications, produces server code, and refines that code through an integrated feedback loop that analyzes execution logs and error messages. By enabling local code execution, our system directly addresses common real-world challenges such as configuration mismatches and environment-specific bugs, ensuring the generated services are functional.

In this study, we extended the findings of our previous work \cite{enase25} by using the PRAB benchmark to expand our evaluation process. The results indicate that the proposed multi-agent system achieved 100\% accuracy in generating OpenAPI specifications from the PRAB benchmark, with the generated results matching the ground truth of seven specifications. For two of the seven specifications, the initial code execution failed. However, after feeding these projects back into the system, it was able to fix the issues and make them executable. Ultimately, code aligned with the specifications was successfully generated for all seven projects.

Moving forward, our next goal is to make the proposed system capable of generating code for larger specifications. This can be achieved by dividing the specifications into smaller, more manageable components, with each component handled individually by the system. Additionally, more functions can be added to the LLM's tool-calling feature to handle complex commands, such as those for Docker, Git, or cloud deployment. This would enable users to interact with the system using natural language prompts, eliminating the need to remember complex command-line instructions.

\bibliographystyle{splncs04}
\bibliography{example}





\end{document}